\documentstyle[12pt,aasms]{article} 
\def\etal{{\it et al.\,}}
\def\flux{{$\times$10$^{-5}$ photons cm$^{-2}$ s$^{-1}$ sr$^{-1}$\,}}

\begin{document}

\title{EGRET Observations of the Extragalactic Gamma Ray Emission}
\author{P. Sreekumar\altaffilmark{1,2},
D. L. Bertsch\altaffilmark{1},
B. L. Dingus\altaffilmark{3}
J. A. Esposito\altaffilmark{1,2}
C. E. Fichtel\altaffilmark{1},
R. C. Hartman\altaffilmark{1},
S. D. Hunter\altaffilmark{1},
G. Kanbach\altaffilmark{4},
D. A. Kniffen\altaffilmark{5},
Y. C. Lin\altaffilmark{6},
H. A. Mayer-Hasselwander\altaffilmark{4},
P. F. Michelson\altaffilmark{6},
C. von Montigny\altaffilmark{7},
A. M\"ucke\altaffilmark{4},
R. Mukherjee\altaffilmark{8},
P. L. Nolan\altaffilmark{6},
M. Pohl\altaffilmark{9},
O. Reimer\altaffilmark{4},
E. Schneid\altaffilmark{10},
J. G. Stacy\altaffilmark{11},
F. W. Stecker\altaffilmark{1},
D. J. Thompson\altaffilmark{1},
T. D. Willis\altaffilmark{6}
}
\altaffiltext{1}{NASA/Goddard Space Flight Center, Code 660,
Greenbelt MD 20771}
\altaffiltext{2}{Universities Space Research Association}
\altaffiltext{3}{University of Utah, Salt Lake City, UT 84112}
\altaffiltext{4}{Max-Planck Institut f\"ur Extraterrestrische Physik,
8046 Garching bei M\"unchen, GERMANY}
\altaffiltext{5}{Hampden-Sydney College, P. O. Box 862, Hampden-Sydney
VA 23943}
\altaffiltext{6}{W. W. Hansen Experimental Physics Laboratory,
Stanford University, Stanford, CA 94305-4085}
\altaffiltext{7}{Landessternwarte Heidelberg-Konigstuhl, Heidelberg, GERMANY}
\altaffiltext{8}{Barnard College, New York, NY 10027}
\altaffiltext{9}{Danish Space Research Institute}
\altaffiltext{10}{Northrop Grumman Aerospace Corporation, Mail Stop A01-26,
Bethpage NY 11714}
\altaffiltext{11}{University of New Hampshire, Durham, NH 03428}

\begin{abstract}
The all-sky survey in high-energy gamma rays (E$>$30 MeV) 
carried out by the Energetic Gamma Ray Experiment Telescope (EGRET)
 aboard the Compton Gamma-Ray Observatory provides a unique  
opportunity to examine in detail the diffuse gamma-ray 
emission. The observed diffuse 
emission has a Galactic component arising from cosmic-ray 
interactions with the local interstellar gas and radiation as 
well an almost uniformly distributed component that is 
generally believed to originate outside the Galaxy. Through a 
careful study and removal of the Galactic diffuse emission, 
the flux, spectrum and uniformity of the extragalactic 
emission is deduced. The analysis indicates that the extragalactic 
emission is well described by a power law photon spectrum 
with an index of --(2.10$\pm$0.03) in the 30 MeV to 100 GeV energy
range. No large scale spatial anisotropy or changes in the energy spectrum 
are observed in the deduced extragalactic emission. 
The most likely explanation for the origin of this 
extragalactic high-energy gamma-ray emission is that it arises 
primarily from unresolved gamma-ray-emitting blazars.
\end{abstract}

\section{Introduction}
Observations at all wavelengths from radio to gamma rays have 
shown the presence of a diffuse background 
emission that appears 
to emanate from beyond our galaxy. It is generally believed 
that these extragalactic photons have different origins, which may be 
broadly thought of as arising from truly diffuse processes or 
from the contributions of a large number of unresolved 
point sources. The X-ray background 
(below 100 keV) is generally understood to arise primarily from
 the integrated emission of unresolved active galactic nuclei (AGN)
 (Leiter and Boldt 1992; Comastri \etal 1995; Zdziarski \etal 1995) such as  
Seyfert galaxies with redshifts ranging up to 5. In the low-energy 
gamma-ray regime (0.5 to 10 MeV) 
the measurement of an extragalactic component is made difficult 
by the presence of significant instrumental background, nuclear line emission and 
emission from strong sources. Recent results from the COMPTEL 
instrument (Kappadath \etal 1996) on the Compton Gamma-Ray 
Observatory (CGRO) and the Solar Maximum Mission (SMM) (Watanabe \etal
1997), indicate the
presence of an extragalactic component, 
significantly lower than previous measurements. 
At higher energies
($>$30 MeV), the OSO-3 satellite provided the first evidence for gamma-ray
emission from our Galaxy (Kraushaar \etal 1972).
This was followed by observations 
by the SAS-2 satellite (Fichtel \etal 1975), which provided clear 
evidence for a strong correlation of the observed diffuse emission
with the expected gamma-ray emission from cosmic-ray, matter 
and photon interactions. In 
addition, the SAS-2 results also indicated for the first time, the presence 
of a residual, apparently isotropic emission which was  
interpreted as not being associated with the Galaxy and hence of 
extragalactic origin (Fichtel \etal 1977; Fichtel, Simpson, and 
Thompson 1978). The SAS-2 telescope had been carefully designed and 
tested to have a very low instrumental background (Derdeyn \etal 1972) 
and launched into a near equatorial orbit to minimize the effects of
trapped radiation. This was necessary to carry out observations of 
this diffuse radiation. In the final analysis of 
the SAS-2 data (Thompson and Fichtel 1982), the integral 
intensity above 100 MeV was determined to be 
(1.3$\pm$0.5)\flux and the slope of the spectrum 
to be --2.35$^{+0.4}_{-0.3}$. Detailed measurements of the isotropy were 
not possible, but the SAS-2 results did show that the 
center-to-anticenter ratio for the radiation within 
20\deg$<|b|<$40\deg was 1.10$\pm$0.19 and the ratio of the 
intensity perpendicular to the plane to that in the 20\deg$<|b|<$40\deg 
region was 0.87$\pm$0.09 (Fichtel, Simpson, and Thompson, 1978). 
Thus, these early results did suggest that there was no evidence for significant
large scale anisotropy in the extragalactic gamma-ray emission. 

With more than an order of magnitude increase in sensitivity over 
previous experiments, a low instrumental background, 
and a low altitude orbit,
EGRET (Energetic Gamma-Ray Experiment Telescope) has provided an 
opportunity to study the spectrum and distribution of the 
extragalactic emission in greater detail than was possible in the past. 
The Compton Gamma-Ray Observatory was launched on  1991 April. A
prime objective of EGRET was to carry out the first all-sky survey 
in high-energy gamma rays (30 MeV$\le$ E $\le$ 30 GeV).
 The most striking feature of the diffuse radiation is the dominant 
emission from the Galaxy which, fortunately for the study of the extragalactic
emission, is strongly concentrated along the Galactic plane. 
The survey also shows the presence of many point-like sources, 
some of which have been identified with 
pulsars, molecular clouds, active 
galactic nuclei, and a nearby normal galaxy (LMC). 
About 51 sources have been identified with a class of 
active galactic nuclei (AGN) called blazars 
(Thompson \etal 1996; Mattox \etal 1997; Mukherjee \etal 1997) which are, 
in general, radio-bright AGN with flat radio spectra, strong optical polarization, 
superluminal motion (often) and significant time variability 
at most wavelengths. EGRET observations have shown that for a majority 
of the gamma-ray luminous blazars, the gamma-ray emission dominates 
the bolometric luminosity (von Montigny \etal 1995), at least during
outbursts. 

This paper discusses the findings on the diffuse 
high-energy gamma-ray emission at high latitudes ($|b|\ge$ 10$^\circ$). 
A detailed description of the intense Galactic plane emission 
is provided by Hunter \etal 1997. We discuss 
the approach used to study the residual, presumably extragalactic high-energy 
gamma radiation by carefully accounting for the foreground 
Galactic emission. The spectral and spatial distributions are determined.
Possible origins of the extragalactic emission are considered, including the 
question of whether it is truly diffuse or resulting 
from unresolved point source contributions.

\section{Observations}

The EGRET instrument is sensitive to gamma rays in the energy 
range from about 30 
MeV to 100 GeV. Gamma rays are detected using a multilevel 
thin plate spark chamber system using the pair production process. 
An internal triggering telescope detects the presence 
of the electron-positron pair with the correct direction of motion and a large 
NaI(Tl) crystal acts as energy-measuring calorimeter. An 
anticoincidence dome provides discrimination against 
the relatively intense charged-particle radiation in space. A 
detailed description of the instrument is given by Kanbach \etal 
(1988) and details on the instrument calibration and analysis 
procedures are available in Thompson \etal (1993), 
Fichtel \etal (1993), and  Esposito \etal (1997). In the 
wide field of view operational mode used for most of the observations 
discussed here, the effective area of the telescope is about 1000 
cm$^2$ at 100 MeV and rises to 1500 cm$^2$ around 0.5 -- 1 GeV, 
and decreases gradually at higher energies to about 700 cm$^2$ 
at 10 GeV near the center of the field-of-view. The effective 
area is a maximum when the target is on axis, falls to 
approximately 50\% of this value when the angular offset reaches 
18\deg, and has a useful sensitivity to at least 30\deg. The 
large field-of-view of EGRET is important for the study of the 
diffuse radiation being discussed here. 

EGRET observations from 1991 April to 1994 Sept. (phases I,II, and III) 
are used in the analysis reported here. These were obtained 
using the full field-of-view mode of the instrument and running the calorimeter 
in coincidence with spark chamber triggers. Maps of counts, exposure, and 
intensity are generated in sky coordinates with 0.5\deg$\times$0.5\deg\, bins
for the individual observations. They are combined to create all-sky maps. 
Only data from within 30\deg of the pointing direction of 
the individual observations, within which the exposure and 
instrument characteristics are best understood, are used in 
creating the added maps. The combined exposure map (see Figure 2. in Thompson \etal
1996) indicates non-uniform sky coverage
which needs to be taken into account during the analysis.

\section{Background Contributions}
Determining the extragalactic diffuse radiation is one of 
the more difficult measurements, because of
 the very low intensity of the expected emission and the lack 
of a spatial or temporal signature to separate the cosmic 
signal from other radiation. As will be shown, the spectrum, 
is significantly different from the Galactic diffuse radiation.
In order to study the extragalactic emission, one must subtract from 
the observation any instrumental background, the Galactic 
diffuse emission, and the contribution from resolved point sources.

\subsection{Instrumental Background}

It is essential to demonstrate that the instrumental background is kept well 
below the level of diffuse emission under study.
Hence three separate factors must be considered in the design and implementation of 
any space-borne high-energy gamma-ray detector used to study the extragalactic diffuse emission: a low cosmic-ray intensity environment for the spacecraft, the detailed 
instrument design, and its placement on the spacecraft.
Only the SAS-2 and EGRET 
instruments had a background well below the intensity of the now 
measured diffuse extragalactic emission. The Compton Gamma-Ray 
Observatory was placed into a standard 28$^\circ$ inclination
orbit using a shuttle launch. For the first 6 years after launch, the
Compton Observatory was kept below $\sim$ 400--450 kilometer altitude 
to minimize the effects of the trapped radiation. 
Important features of the instrument design include the directional trigger system, 
the highly efficient (1 part in 10$^6$) anticoincidence dome, 
and the amount and placement of the material above the anticoincidence 
dome. The typical cosmic-ray flux in orbit that EGRET sees, is about 
10$^4$ times the average gamma-ray flux.
The high efficiency for rejection of cosmic-ray 
events by the anticoincidence dome ensures that nearly all cosmic-ray 
induced events are removed. However, cosmic rays impinging 
on material above the anticoincidence dome could produce,
neutral pions via inelastic collisions which decay into 
gamma rays. Clearly, the material must be kept to a minimum consistent 
with thermal and light penetration, but it must also be kept very 
close to the anticoincidence scintillator to maximize the 
probability that a secondary in any of the 
very improbable interactions will penetrate and trigger the 
anticoincidence dome. A new problem arose in the design of 
EGRET late in its development when, during the repair of the Solar 
Maximum Mission satellite, it was discovered that the space debris 
had become much more intense than previous estimates. A 
penetration of the light shield surrounding 
the anticoincidence dome would cause a small light leak, seriously degrading 
its performance significantly. Through a special design of the covering material 
the total amount of matter on the outside of the dome was held on an
average to only 
20$\%$ more than that of SAS-2 ($\sim$ 0.19 g\,cm$^{-2}$ compared 
to $\sim$ 0.16 g\,cm$^{-2}$). In addition, this material was kept close 
to the anticoincidence dome in order to have the highest probability for a 
secondary charged-particle to enter and trigger the anticoincidence 
system. 
Extensive preflight calibration was carried out at the Brookhaven Laboratory
to examine the background contribution from the material above the
anticoincidence dome. These tests demonstrated that the 
instrumental background was more than an
order of magnitude less than the intensity of the extragalactic diffuse radiation 
derived from the SAS-2 data (Thompson \etal 1993). 
Furthermore, a steeper spectrum is expected for the cosmic-ray induced 
instrumental background.

A potential source of contamination is from Earth albedo gamma rays. These
gamma rays arise from cosmic-ray interactions in the Earth's upper 
atmosphere and are indistinguishable from cosmic gamma rays in the spark chamber. 
EGRET triggering is disabled during the Earth 
occultation part of the orbit. As described in Thompson \etal (1993),
EGRET carries two 4 by 4 arrays of scintillator tiles which allow a
set of 96 possible sub-telescope combinations. As the Earth enters the
field of view of any of the direction modes, triggering is disabled
for that mode and thus events with an arrival direction defined 
by that direction mode, are not accepted.  However, the albedo gamma-ray flux 
is quite large and peaks towards the horizon where it is about 10 
times more intense than towards the nadir (Thompson, Simpson, 
and \"Ozel, 1981), resulting in the inclusion of a large number of Earth 
albedo events in the primary data. 
The elimination of these Earth albedo events is carried out as follows. 
An energy dependent cutoff in zenith angle ($\phi_{cutoff}(E)$) 
is defined where the zenith angle ($\phi(E)$ is defined as the angle made by the 
incoming gamma ray with the direction of the Earth center 
to spacecraft vector. Events with zenith angles larger than 
$\phi_{cutoff}(E)$ are rejected, where

{$\phi_{cutoff}(E) = 110\deg - \sigma\theta(E)$} \hspace{1.5cm}
where  {$\theta(E)=5.85\deg (\frac{E}{100MeV})^{-0.534}$}
is the calibrated instrument point spread function (Thompson \etal
1993).
In order to sufficiently eliminate the inclusion of albedo events, 
a $\sigma$ value of 4.0 is used for the analysis of the 
extragalactic emission, which is more restrictive than the value 
2.5 used for the standard EGRET analysis of point sources. 
The value of $\phi_{cutoff}(E)$ is further restricted to a 
minimum value of 80\deg at low energies to minimize excessive loss of
good events and not to exceed a maximum 
of 105\deg at high energies.  Use of the new values of $\phi_{cutoff}(E)$ 
also removed some albedo events that appeared to show non-uniform
spatial clustering in the summed data (Willis 1996). Since there is 
only a negligible
loss of good events as a result of this tighter restriction in zenith
angle, no corrections are applied to the data.

A final check of the contamination of the in-flight data 
from the charged-particle background
is carried out by studying the rigidity dependence of the accepted events. 
The cut-off rigidity for charged cosmic rays, which varies throughout the orbit, 
defines the minimum momentum per unit charge necessary to avoid being 
excluded by the Earth's magnetic field. So regions along the orbit 
characterized by high-magnetic field strengths have a smaller cosmic ray flux. 
Figure 1 shows a rigidity plot of the histogram of event
rates after imposing all selection criteria, for a range of 
observations. A KS-test using rigidity intervals of $<$0.1 GV yielded
negligible probability for the distribution of event rate to be
correlated with rigidity, implying there is no significant charged-particle 
contribution to the extragalactic emission being discussed here. 
There is indeed the possibility of high-rigidity ($>$20GV) cosmic ray particles 
contributing to the background since they are not modulated by the
Earth's magnetic field.  However, these charged particles have a high 
probability of producing additional charged particles along with the
gamma rays during the interaction with the material ahead of the 
anticoincidence dome and thus enable the anticoincidence system 
to reject the event.

\subsection{Galactic diffuse emission}
The Galaxy is a strong source of high-energy gamma rays arising primarily from 
the interaction of cosmic rays with the interstellar matter and to a 
lesser extent, interstellar photons.  This emission is strongest within 
$\pm$60\deg\, in Galactic longitude and $\pm$10\deg\, in Galactic latitude,
where most of the interstellar gas is present. It falls off 
rapidly at higher latitudes. The primary processes that 
produce the observed Galactic diffuse gamma rays are: cosmic-ray nucleons 
interacting with nucleons in the interstellar gas, bremsstrahlung by 
cosmic-ray electron, and inverse Compton interaction of cosmic-ray electrons
 with ambient low-energy interstellar photons (Bertsch \etal 1993; Hunter \etal 1997). 
The possible contribution from unresolved point sources such as pulsars, 
is uncertain with estimates ranging from a few percent to almost 100$\%$ depending on
the birth properties of pulsars (Bailes and Kniffen 1992). The evidence for a pion
"bump" in the Galactic diffuse spectrum (Hunter \etal 1997) implies the unresolved
source contribution is well below 50$\%$; however a contribution $\le$ 20$\%$ cannot 
be ruled out at present. No effort is made to incorporate an unresolved source 
component in the Galactic diffuse calculations used in this study.
The electron bremsstrahlung and neutral pion decay
processes dominate the diffuse emission, although at higher latitudes the 
inverse Compton process could contribute up to 30\% of the total observed 
Galactic radiation, particularly in the inner Galaxy. The calculations
 use deconvolved matter density distribution along with some assumptions about the 
corresponding cosmic-ray density distribution. Bertsch \etal (1993) 
and Hunter \etal (1997) discuss the detailed 
calculation of the diffuse emission from the Galactic plane, including 
the approach used to derive the Galactic cosmic ray density distribution. 
At latitudes $|b|\ge$ 10$^\circ$, the calculation of 
the Galactic diffuse emission is more straightforward, since the 
gas distribution along the line-of-sight is mostly local (within a 
few hundred parsec) and the corresponding cosmic-ray spectrum at energies important for
gamma-ray production is not expected to be 
very different from that which is measured locally at Earth, corrected
for solar modulation. 

The primary inputs necessary to calculate the expected level of diffuse
gamma-ray emission in our galaxy include the distribution of interstellar hydrogen gas
(neutral and ionized), low-energy photon distribution (3K, IR, visible, UV), 
and the cosmic-ray density, spectrum and composition in the Galaxy 
along every line-of-sight.  As stated before, the high-latitude 
($|b|\ge$ 10$^\circ$) calculation assumes, the spectrum of
cosmic rays (electrons and protons) is the same as that 
measured locally, corrected for solar modulation. The neutral atomic hydrogen 
distribution is derived from the 21-cm map of Dickey and Lockman (1990) 
while the molecular 
hydrogen distribution is derived indirectly using the strength of 
$^{12}$C$^{16}$O emission at 2.6 mm provided by Dame \etal (1987) 
(converted into H$_2$ column density including a normalization factor of 1.5 
$\times 10^{20}$cm$^{-2}$(K km s$^{-1}$)$^{-1}$(Hunter \etal 1997)). Recent enhancements to the Dame \etal 
work by Digel (1996) are included. The ionized 
interstellar medium is not as well determined.  The model of Taylor 
and Cordes (1993) based on radio pulsar studies is used here. 
At intermediate latitudes (10\deg$< |b|\le$30\deg),
most of the Galactic emission arises from cosmic ray interactions with neutral 
material, while at higher latitudes ($>$30\deg) the contributions 
from the ionized medium and the inverse Compton component become 
important due to the larger scale heights of the ionized gas and
the low-energy photons. Since the spatial 
distribution of the low-energy photon density  
(NIR, FIR, Optical and UV) in the Galaxy is uncertain, 
the inverse Compton calculation is not well determined. 
Furthermore, the true cosmic-ray electron scale height is not known. 
The deconvolution of the observed Galactic radio synchrotron
data provides only a lower limit to the cosmic-ray electron 
scale height of 1 kpc (Broadbent, Haslam and Osborne 1989) since the Galactic magnetic field 
distribution is not well understood. Consequently, the current 
results on the extragalactic background emission are subject 
to these uncertainties, but recall that this is not the dominant
component. Hunter \etal (1997) have shown that there is good agreement
between the observed high-energy gamma-ray diffuse radiation in the
Galactic plane and the calculations. In the next section, the 
calculations are compared with 
observations at all latitudes. The method for treating uncertainties
in deriving the extragalactic emission is also discussed.

\section{Analysis}

The contribution from the 157 sources from the 2nd EGRET source catalog
and its supplement (Thompson \etal 1995, 1996), observed in excess of the 
diffuse background was modeled and subtracted from the data. A new set of 
binned maps (0.5\deg $\times$ 
0.5\deg pixels) of event counts and instrument exposure 
containing only contributions from 
unresolved sources, Galactic diffuse \& extragalactic diffuse radiation,
were used for the analysis that follows.
Since the determination of the residual, presumably extragalactic radiation, 
depends critically on the degree to which the calculated Galactic 
component matches the true Galactic diffuse emission, it is important 
to demonstrate here the consistency of the predictions with the observations 
at all latitudes. 
Figure 2 shows calculated and observed
emission profiles in Galactic latitude averaged over 60\deg bands of Galactic longitude.
These results indicate the high-latitude calculations are in
good agreement in general with the observed diffuse emission. However, in the inner Galaxy near the Galactic center 
($\sim |l|\le 40\deg \& |b|\le 30\deg$) the calculated intensities fall below 
the observed values. This could arise from interstellar material that 
is not included in the model or additional contributions from inverse 
Compton processes arising from unmodeled soft photon distributions or
unresolved sources.
In addition, the local cosmic-ray density distribution could be non-uniform.

In order to derive the extragalactic emission without being very sensitive 
to the Galactic model used, the following approach is adopted.
As described in Section 1, the observed emission at high latitudes 
is assumed to be made up of a Galactic and an extragalactic component. 

\centerline{$I_{total}(l,b,E) = A + B \times I_{Gal}(l,b,E)$}

\noindent where $I_{total}(l,b,E)$ is the total diffuse intensity of high-energy gamma
radiation as a function of Galactic coordinates, and the gamma-ray
energy E, and $I_{Gal}(l,b,E)$ intensity of the Galactic diffuse emission.
The slope, "B" of a straight line fit to a plot of observed emission versus 
the Galactic model gives an independent measure for the normalization of
the model while the intercept,"A", equals the extragalactic emission. 
Figure 3 shows the correlation plot for the 10 standard EGRET energy
intervals, and Table 1 summarizes the results. The points represent 
the average intensity within areas of 
equal solid angle ($\sim$ 0.1 steradian) covering the whole sky except
for an exclusion zone about the Galactic plane ($|b|< 10\deg$) and the
region around the Galactic center ($|l|\le 40\deg \, \& \, 10\deg\le 
|b|\le 30\deg$).
The choice of minimum solid angle bins was made after taking into 
consideration the limited counting statistics at high and low energies 
and the distribution of the non-uniform sky coverage. 
The correlation is not as good at the lowest energy interval of 30 --
50 MeV and at the highest energy interval (4--10 GeV), primarily due to
limited counting statistics. At intermediate energies, the strong correlation
provides tight constraints on the normalization of the Galactic diffuse
emission. The rather significant deviations in the normalization
parameter from unity, for energies $>$ 1 GeV, reflects the spectral 
discrepancy between the Galactic diffuse model and observations 
and is discussed in detail by Hunter \etal (1997). This approach yielded
the extragalactic gamma-ray spectrum up to 10 GeV. Beyond 10 GeV 
larger uncertainties exist with the EGRET sensitivity calculation
due to the self-veto arising from back-splash in the NaI calorimeter
(Thompson \etal 1993). Hence, the results of a Monte Carlo simulation 
(Lin 1992) are used to determine the differential flux 
in the 10--30, 30--50, and 50--120 GeV energy ranges. Consequently,
these spectral points have additional associated systematic uncertainties.
In this analysis, a systematic error of 13$\%$ (see Esposito \etal 1997)
is assumed for 30--10000 MeV and 30$\%$ beyond 10 GeV. This also takes into
account errors in correction factors used to model the changing spark
chamber performance with time and photon energy.

In order to examine the isotropy of this emission, the sky was 
divided into 36 independent regions which excluded the region $|b|< 10\deg$ 
at all longitudes as well as $|l|\le 40\deg \, \& \, 10\deg\le |b|\le 30\deg$ 
about the Galactic Center. The latitude regions near the Galactic 
center are excluded due to difficulties in accounting for all 
the Galactic diffuse emission, while the Galactic plane is excluded 
due to the extreme dominance of the observed emission by the 
Galactic component. Finally, in deriving the spectrum of the 
extragalactic emission in each of these regions, it was assumed 
that the energy-dependent normalization factors for the Galactic calculation 
(Table 1), derived from the all-sky analysis, remains unchanged for the 36 independent
regions of the sky. It would be more appropriate to determine the
normalization factors separately in each of the individual regions on the sky;
however, statistical limitations give large uncertainties in the results of the
correlation analysis.

\section{Results}
Using the approach described in Section 4, the resulting differential photon 
spectrum of 
the extragalactic emission averaged over the sky is well fit by a power 
law with an index of --(2.10$\pm$0.03). The spectrum was determined using 
data from 30 MeV to 10 GeV; however as shown in Figure 4, the differential 
photon flux in the 10 to 20 GeV, 20 to 50 GeV, and 50 to 120 GeV energy 
intervals are also consistent with extrapolation of the single power law spectrum. 
The integrated flux 
from 30 to 100 MeV is (4.26$\pm$0.14)\flux and that above 100 MeV is 
(1.45$\pm$0.05)\flux were determined from the intercept of the straight
line fit to the total observed emission and the Galactic model. 
This agrees with previously reported SAS-2 results (Thompson and Fichtel 1982), however the spectrum is slightly harder.

EGRET observations also permit an examination of the degree of isotropy of the 
extragalactic emission. The derived integral fluxes 
in the 36 independent regions of the sky are shown in Figure 5. 
The flux values range from a low of (0.89$\pm$0.17)\flux to a high of (2.28$\pm$0.34)\flux.
Figure 6a shows a histogram of integral flux values, overlayed by a
gaussian fit assuming equal measurement errors. 
The gaussian fit yields a mean flux above 100 MeV of 1.47\flux with a 
1$\sigma$ value of 0.33\flux consistent with the all-sky calculation. 
Of the 36 measurements, 69$\%$ fall within 1$\sigma$ of the mean, 97$\%$ fall within
2$\sigma$, with only the largest value falling outside the 95$\%$ confidence
interval (Figure 6a). 
To further examine the spatial distribution, Figure 6b and
6c shows the same data plotted against Galactic longitude and 
latitude respectively.
No significant deviation from uniformity
is observed in the latitude distribution; however as a function of longitude, 
there appears to be an enhancement in the derived intensities 
towards $\sim$l=20\deg, but a closer examination shows this feature to be due to a few
regions on the boundary of the excluded Galactic center region. 
This could arise from unaccounted Galactic diffuse emission.  
If one excludes the inner regions of the Galaxy ($|$r$|$ $<$60\deg;
r=angle made with the direction of the Galactic center), the 
mean flux is 1.36\flux above 100 MeV.
The distribution of the 28 independent observations is consistent with
isotropy (reduced $\chi^2$ = 1.09).
As discussed below, the inclusion of additional Galactic emission in the
inner Galaxy measurements is not inconsistent with the observed
distribution of spectral indices.
However, the exact nature of the `excess' emission in the inner
Galaxy is not resolved by this analysis.

Figure 7 shows the distribution of spectral index in 36 independent 
regions of the sky. The power law indices vary from --(2.04$\pm$0.08) to --(2.20$\pm$0.07) and show no systematic 
deviations from the value of --(2.10$\pm$0.03), derived from the all-sky analysis.
The histogram of spectral indices shown in Figure 8a yields a mean value
of --2.11 with a 1$\sigma$ value of 0.04. The corresponding longitude and
latitude distributions (Figure 8b, 8c) show no indication of a smooth variation in the index. 
The expected distribution of the spectral indices  can be calculated if 
one assumes that the outer Galaxy average represents a better measure 
of the isotropic extragalactic emission and
that the `excess' emission near the Galactic center region arises solely
from unaccounted Galactic diffuse emission. The expected distribution should
be characterized by an average of --2.12 with a standard deviation of 0.04.
This is consistent with  the average value of --(2.11$\pm$0.04) derived from the all-sky
distribution shown in figure 7.
Finally, a distribution of the integral flux above 100 MeV versus the 
corresponding spectral index is shown in Figure 9.
Even though the Galactic diffuse emission spectrum is not well characterized
by a single power law (Hunter \etal 1997), such a characterization in the 30--10000 MeV
energy range would
in general yield a harder spectral index($\sim$1.8--1.9) than that of the derived
extragalactic emission. Hence, any significant Galactic 
contribution to the derived extragalactic emission would show evidence 
for some correlation between the integral flux and the spectrum.
With the exception of a few points with the highest intensity, there is
no correlation between the intensity and the spectral index. As stated
before, these high points by themselves only suggest the presence of a 
small residual Galactic component in some of the inner Galaxy measurements.

\section{Discussion}
The results just presented have confirmed the existence of a generally 
uniform, presumably extragalactic diffuse radiation, consistent in intensity 
with that determined by the SAS-2 experiment. They are generally consistent with the
preliminary results from Kniffen \etal (1996), Osborne,
Wolfendale and Zhang (1994), and Chen, Dwyer, and Kaaret (1996) which
were all based on a smaller set of observational data.  
The presence of significant `excess' emission from (l$\sim$90\deg,b$\sim$52\deg)
to (l$\sim$45\deg,b$\sim$77\deg) reported by Chen, Dwyer, and Kaaret (1996),
is not as evident in our analysis possibly due to the more stringent
zenith angle selection.
Osborne, Wolfendale and Zhang obtained a smaller integral
flux, probably resulting from attributing a larger contribution from the
inverse Compton process to the Galactic diffuse emission. 
The results presented here using a larger dataset
and a much more careful accounting of other contributing effects, have greatly 
extended the energy spectral information, both in precision and energy range, as well as 
provided substantial additional information on its uniformity.  Combined with the x-ray 
and low-energy gamma-ray information a spectrum for the uniform diffuse radiation now 
exists over a broad energy range (Figure 10). 

A large number of possible origins for the extragalactic diffuse gamma-ray 
emission have been proposed over the years.  
Fichtel and Trombka (1981) have reviewed some of the proposed models.
Theories of diffuse origin include scenarios of
baryon symmetric universe (Stecker, Morgan, and Bredekamp, 1971), 
primordial black hole evaporation (Page and Hawking, 1976; Hawking, 1977), 
million solar-mass black holes which collapsed at high redshift (z$\sim$100) 
(Gnedin and Ostriker 1992), and some exotic source proposals, 
such as annihilation of supersymmetric particles
(Silk and Srednicki 1984, Rudaz and Stecker 1991, Kamionkowski, 1995).
All of these theories predict continuum or line contributions 
that are unobservable above 30 MeV with current instruments.

Models based on discrete source contributions have considered a variety of source classes.
Normal galaxies might at first appear to be a reasonable possibility for the origin of the 
diffuse radiation since they are known to emit gamma rays and to do so to very high 
gamma-ray energies (Sreekumar \etal 1992; Hunter \etal 1997).  Previous estimates 
(Kraushaar \etal 1972; Strong \etal 1976; Lichti, Bignami, and Paul, 1978; 
Fichtel, Simpson, and Thompson, 1978) have shown
that the intensity above 100 MeV expected from normal galaxies is only 
about 3$\%$ to 10$\%$ of what is observed.  Further and perhaps even more significant, the 
energy spectrum of the Galactic diffuse gamma rays is significantly different from that 
measured by EGRET for the extragalactic diffuse radiation, being harder at low energies
(100 to 1000 MeV) and considerably steeper in the 1 to 50 GeV region.  
Dar and Shaviv (1995) have suggested that emission arises from cosmic ray
interaction with intergalactic gas in groups and clusters of galaxies.  Although 
the authors claim that this proposed explanation leads to a higher intensity level, 
it is still in marked disagreement with the measured energy spectrum (Stecker and Salamon 1996a).
Clearly, small contributions from a number of these proposed sources remains viable.
A fluctuation analysis (Willis 1996) indicates that $\sim$10$\%$ to 100$\%$ 
of the emission could be made up of
unresolved point sources and does not very well constrain the likely contribution from a
diffuse origin.

It has been postulated for over two decades by a large number of authors that active 
galaxies might be the source of this general high-energy gamma-ray diffuse emission (e.g. Bignami
\etal 1979; Kazanas and Protheroe 1983).  
However, prior to the results from EGRET, only one quasar  had been seen in high-energy 
gamma rays (3C273); so there was little data upon which to base a calculation of the 
expected intensity.  Now, a large number of high-energy gamma-ray-emitting blazars have 
been observed by EGRET (von Montigny \etal 1995; Mukherjee \etal 1997).   
One of the more important pieces of evidence in favor of the blazar origin of the high-
energy portion of the diffuse spectrum is the spectrum.  Both the spectrum reported here 
and the average spectrum of blazars may be well represented by a power law in photon 
energy.  The spectral index determined here for the diffuse radiation is --(2.10$\pm$0.03), and 
the average spectral index of the observed blazars is --(2.15$\pm$0.04).  These two numbers 
are clearly in good agreement. A standard cosmological 
integration of a power law in energy yields the same functional form and slope. Considering
the new, well determined gamma-ray spectrum, this argues strongly
that the bulk of the observed extragalactic gamma-ray emission can be explained as
orginating from unresolved blazars.

In order to estimate the intensity of the diffuse radiation from blazars, knowledge of the 
evolution function is needed, as well as the intensity distribution of the blazars.  
Two approaches have been utilized to determine the gamma-ray evolution.
One way is to 
assume that the evolution is similar to that at other wavelengths.  The other alternative is to 
deduce the evolution from the gamma-ray data itself and hence have a solution that depends 
only on the gamma-ray results.  The advantage of the former is that, if the assumption of a 
common evolution is correct, an estimate with less uncertainty is obtained.  The positive 
aspect of the latter is that there is no assumption of this kind, but the uncertainty in the 
calculated results is relatively large because of the small gamma-ray  blazar sample.

Several authors have estimated the contribution from blazars using the first approach
described above (Padovani, \etal 1993;
Stecker, Salamon and Malkan, 1993;
Setti and Woltjer, 1994; Erlykin and Wolfendale, 1995; Stecker and
Salamon 1996b) where one accepts the proposition that the evolution
function determined from the radio data may be applied to
the gamma-ray-emitting blazars. 
The recent work of Mukherjee \etal (1997) shows that there is agreement within
uncertainties between the radio and high-energy gamma-ray z distributions of 
both types of blazars, i.e.  flat-spectrum radio quasars and BL Lacs.
The calculations agree with the observation of the intensity
of the diffuse radiation, with the uncertainty varying with the approach and 
estimated errors.  In most cases, the calculations are within 50$\%$ of the 
observed spectrum.  A word of caution seems appropriate, since it should be
pointed out that the degree and nature of the correlation between the radio and
gamma-ray emission is still being debated (Padovani \etal 1993;
M\"ucke \etal 1996; Mattox \etal 1997).
The calculations of Stecker and Salamon (1996b) predict a curvature in the blazar 
background spectrum; our data while being well represented by a single power law, 
do not rule out such a curvature at least up to few GeV.

Chiang et al. (1995) used only the gamma-ray blazar data 
to deduce the evolution 
function for the gamma-ray-emitting blazars.  They  used the V/Vmax approach in the 
context of pure luminosity evolution (currently the preferred concept) to show that there 
was indeed evolution of the high-energy gamma-ray emitting blazars and to deduce the 
parameters describing the evolution.  These authors found that the evolution was consistent 
within errors to that seen at other wavelengths.  Their estimated intensity 
level ranges from (1.9$^{+8.5}_{-1.4}$ to 2.6$^{+11.4}_{-1.8}$)$\times$10$^{-5}$
photons cm$^{-2}$ s$^{-1}$ sr$^{-1}$ for E$>$100 MeV indicate large 
uncertainties due to the unconstrained nature of the lower end of the 
luminosity function. 

With regard to the isotropy, as described earlier, 
the data in the region $\sim$60\deg away from the Galactic center,
show no detectable deviation from 
isotropy within the limits of the EGRET measurements. The slight
enhancement in the derived extragalactic intensity towards the
Galactic center is believed to arise mostly from additional Galactic
inverse Compton contribution that has not been modeled due to large
intrinsic uncertainties in the soft photon distribution.
No evidence for any smooth angular variation is present in the power law 
spectrum used to characterize the extragalactic emission.

In examining the extragalactic diffuse spectrum from the x-ray to gamma-ray 
spectral region, it has been suggested (Zdziarski 1996) that the combination 
of Seyfert I and Seyfert II and a likely 
contribution from supernovae (The \etal 1993), appears sufficient 
to explain the observed radiation from 1 keV to a few MeV. Although at
this point in time, both the observational and theoretical picture of
the extragalactic gamma-ray emission below 10 MeV is unclear, it appears
that at higher energies, EGRET observations discussed in this paper
strongly suggest that the predominant origin of the gamma radiation in the region 
from 30 MeV to at least 50 GeV is blazar emission.  Other explanations for this 
high-energy gamma-ray region seem to be inconsistent with at least one aspect of the 
observations, with the well measured high-energy gamma-ray spectrum being a particularly 
strong discriminator.  

If the hypothesis that the general diffuse radiation is the sum of the emission of blazars is 
accepted, there is an interesting corollary.  The spectrum of the measured
extragalactic emission implies the average quiescent energy spectra of blazars extend 
to at least 50 GeV and maybe up to 100 GeV without a significant change in slope.  
Most of the measured spectra of 
individual blazars only extend to several GeV and none extend above 10 GeV, simply 
because the intensity is too weak to have a significant number of photons to
measure. Intergalactic absorption does not have much effect at this energy 
except for blazars at relatively large z, and, in any case would have a depressing 
effect on the spectrum.  
Hence, the continuation of the single power law diffuse spectrum 
up to 100 GeV really implies that the source spectrum also 
continues without a major change in spectral slope to at least 100 GeV.  
This conclusion, in turn, implies that the spectrum of the parent 
relativistic particles in blazars that produce the 
gamma rays remains hard to even higher energies.

\acknowledgments
The authors wish thank Andrzej Zdziarski for providing the results of the various
model calculations in machine readable form and Betty Rots for valuable editorial comments. 
The EGRET team gratefully acknowledges support from the following:
Bundesministerium f\"ur Bildung, Wissenschaft, Forschung und Technologie grant 50 QV 9095 (MPE), NASA Cooperative Agreement NCC 5-95 (HSC), NASA Grant NAG5-1605 (SU),
and NASA Contract NAS5-96051 (NGC). 

\clearpage
\begin{planotable}{ccl}
\tablewidth{10cm}
\tablecaption{Results from the all-sky correlation analysis}
\tablenum{1}
\tablehead{
\colhead{Energy (MeV)} &
\colhead{Normalization(B)} &
\colhead{A}    }
\startdata
30--50		&1.14 	&(1.20$\pm$0.35)E-06 \\
50--70		&1.03	&(6.63$\pm$1.29)E-07 \\
70--100		&1.09	&(2.61$\pm$0.35)E-07 \\
100--150	&1.05   &(1.10$\pm$0.15)E-07 \\
150--300	&0.97	&(3.60$\pm$0.48)E-08 \\
300--500	&0.97   &(9.85$\pm$1.34)E-9 \\
500--1000	&1.09   &(2.72$\pm$0.37)E-9 \\
1000--2000	&1.34   &(6.17$\pm$0.84)E-10 \\
2000--4000	&1.85   &(1.52$\pm$0.22)E-10 \\
4000--10000	&1.56	&(3.26$\pm$0.48)E-11 \\
\tablenotetext{}{A = extragalactic flux in photons (cm$^2$-s-sr-MeV)$^{-1}$}
\end{planotable}

\clearpage
\section*{Figure Captions}
\begin{enumerate}
\item Accepted gamma-ray events as a function of geomagnetic rigidity.
The absence of a rigidity dependence,
implies no measurable contamination by charge-particle induced events.
\item Latitude profiles of the observed Galactic emission profile averaged 
over Galactic longitudes (data points). The solid line indicates the calculated
values.
\item Correlation plots of observed diffuse intensity and the calculated Galactic 
emission. The straight line fit provides a normalization factor(B) to the calculated
Galactic model while the intercept(A) gives the residual emission, independent of the
Galactic emission model (see Table 1 for a summary of fit results.).
\item The extragalactic diffuse emission spectrum from 30 MeV to 120 GeV
\item The distribution of extragalactic flux (E$>$100 MeV). The shaded region containing
the Galactic plane is
excluded due to extreme dominance of the Galactic emission and the region centered
on the Galactic Center and extending towards $\pm$30\deg in latitude is excluded due
to difficulties in modeling all of the Galactic emission.
\item The all-sky distribution of integral flux (E$>$100 MeV) in units
of photons (cm$^2$-s-sr)$^{-1}$. (a) Gaussian fit to the flux histogram showing
the mean (solid), 1$\sigma$ (dotted) and 2$\sigma$ (dashed) confidence 
intervals; (b)\&(c) Intensity values plotted over longitude and latitude respectively.
The longitude distribution shows slightly larger intensity values within $\pm$60\deg (see discussion in Section 5).
\item The distribution of power-law spectral indices.
\item The all-sky distribution of power-law spectral index.
(a) Gaussian fit to the spectral index histogram showing
the mean (solid), 1$\sigma$ (dotted) and 2$\sigma$ (dashed) confidence 
intervals. (b)\&(c) Spectral index values plotted over longitude and latitude
respectively, showing the absence of any variation over the sky.
\item Integrated intensities (E$>$100 MeV) plotted against the derived 
power-law spectral indices for different regions of the sky, indicate no
evidence for any correlation.
\item Multiwavelength spectrum of the extragalactic gamma-ray spectrum from X-rays
to high-energy gamma rays. The estimated contribution from Seyfert 1 (dot-dashed), and Seyfert 2
(dashed) are from the model of Zdziarski(1996); steep-spectrum quasar contribution
(dot-dot-dashed) is taken from Chen \etal (1996b); Type Ia supernovae (dot) is from The,
Leising and Clayton (1993). The blazar contribution below 4 MeV (thin long 
dashed) is derived assuming the average blazar spectrum breaks around 4
MeV (McNaron-Brown \etal 1995) to a power law with an index of $\sim$--1.7.
The thick solid line indicates the sum of all the components.

\end{enumerate}
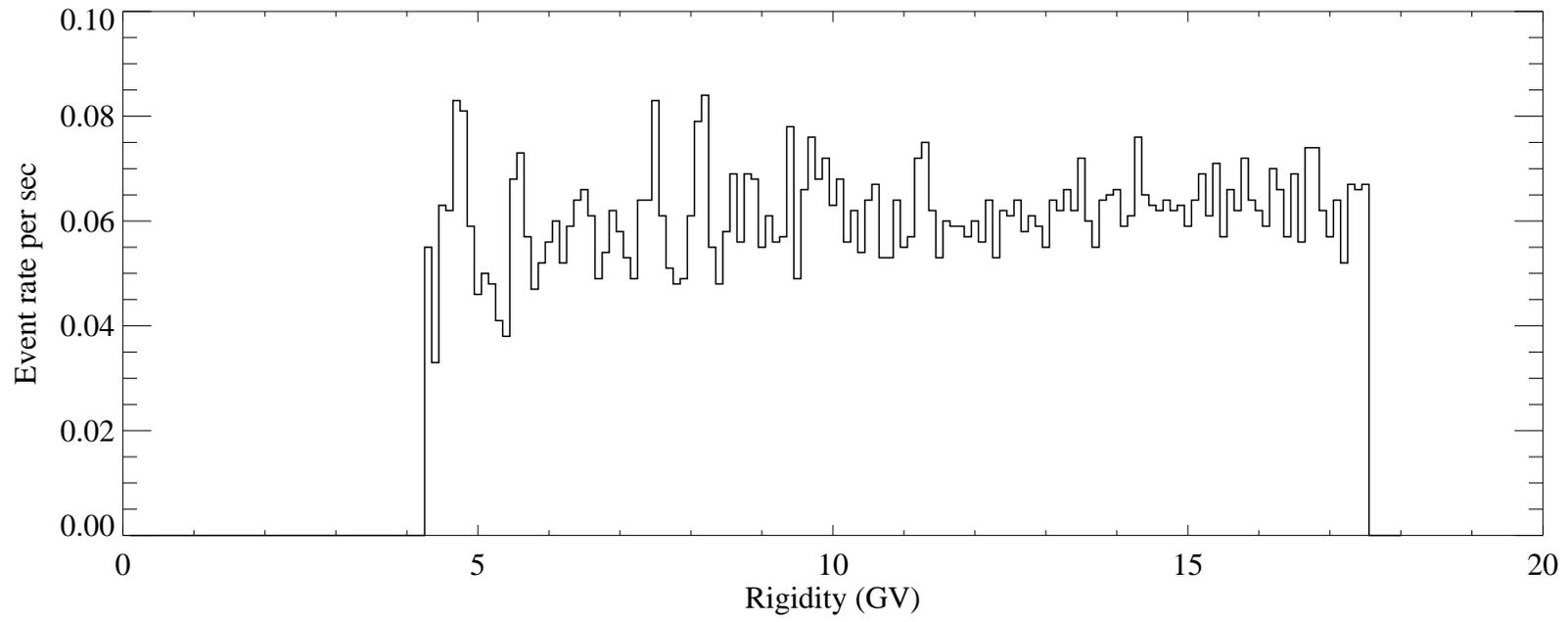
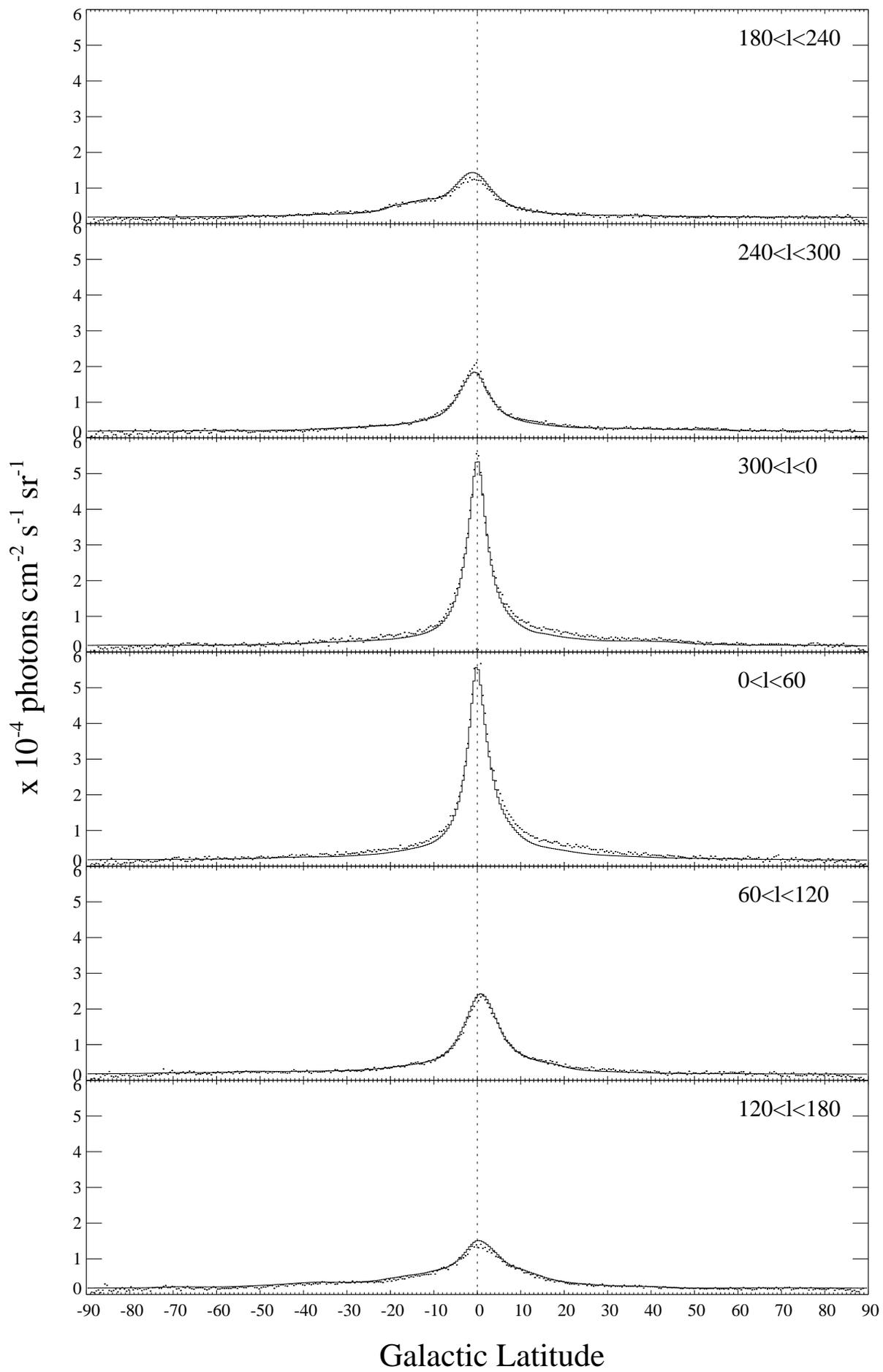
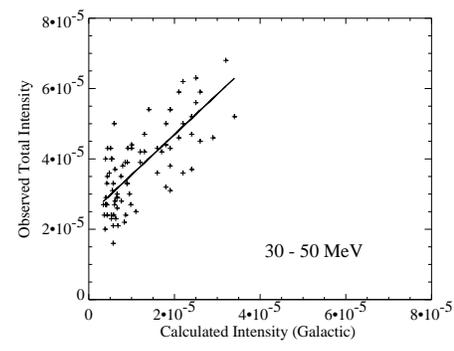
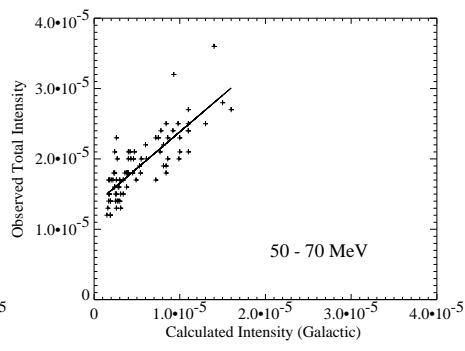
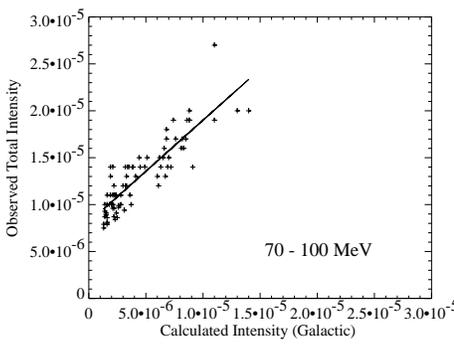
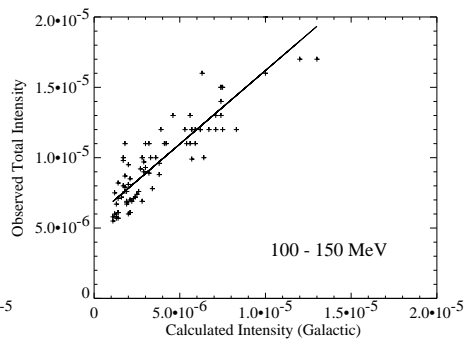
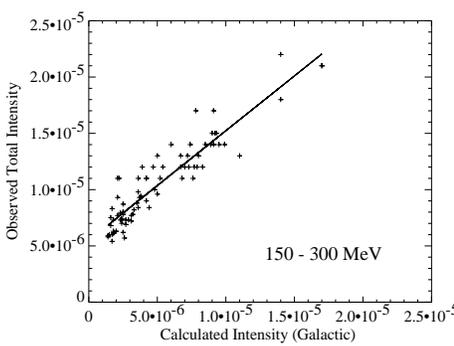
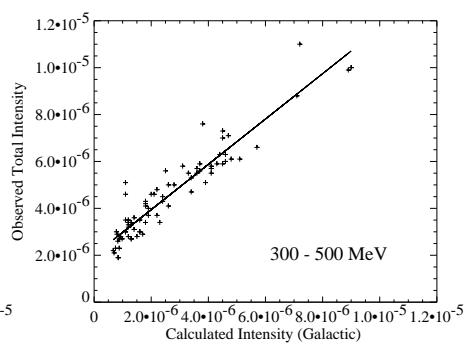
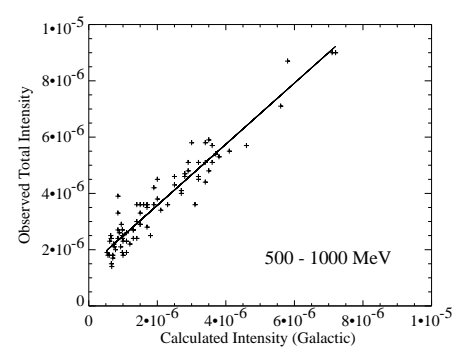
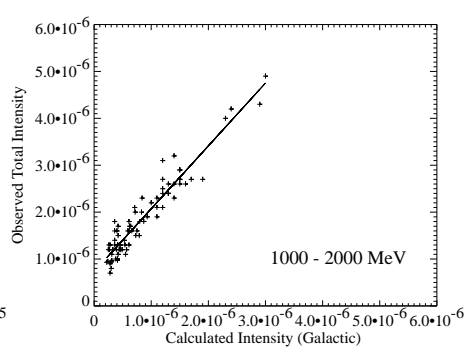
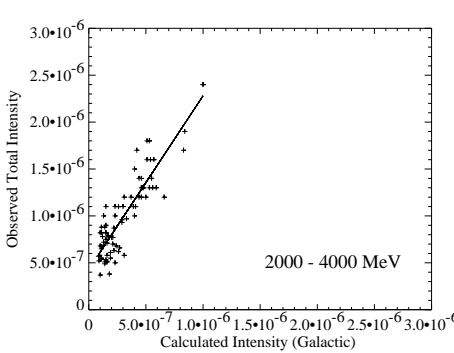
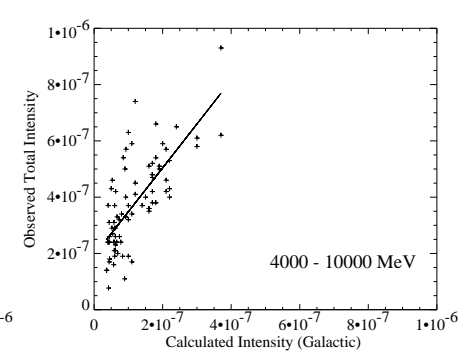
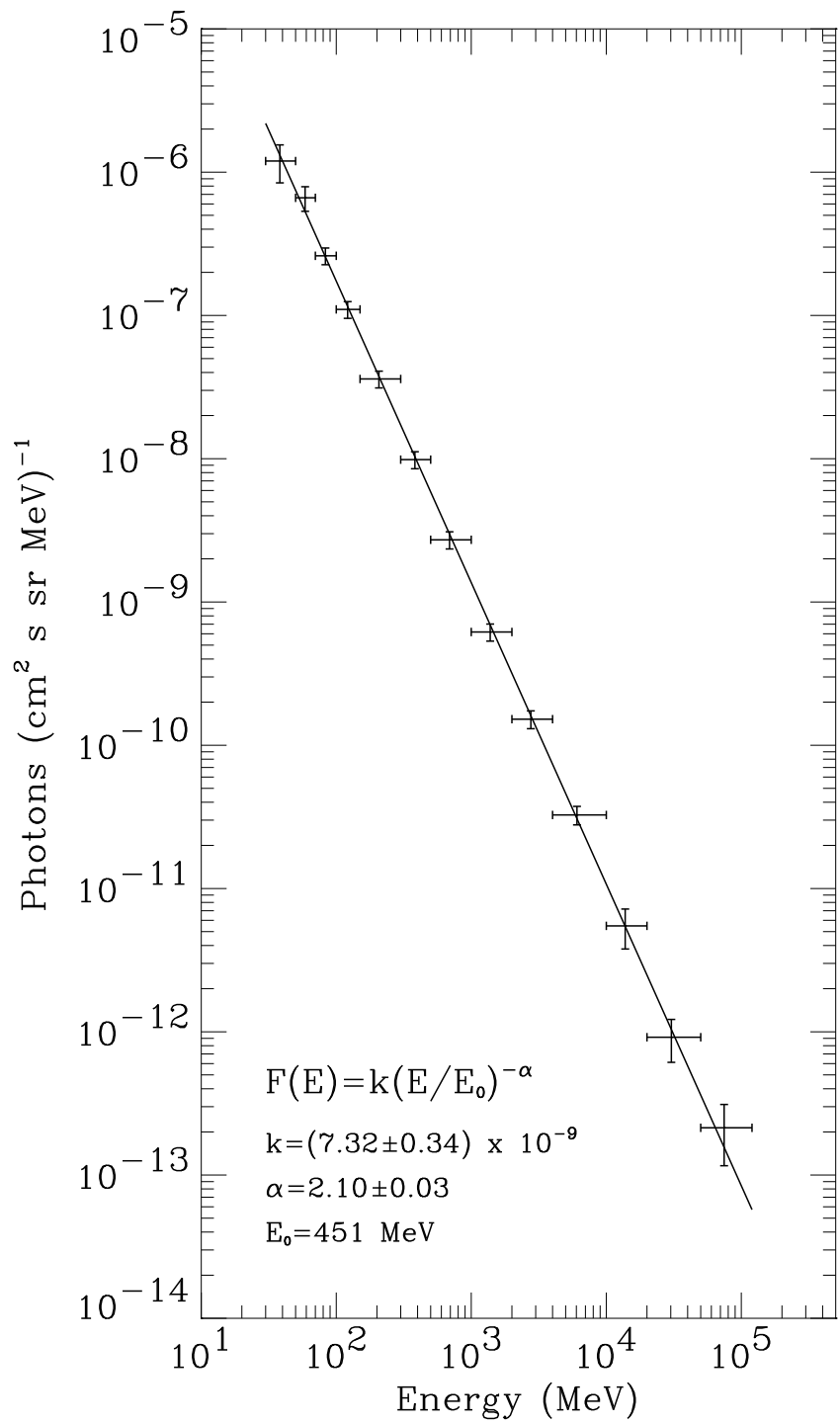
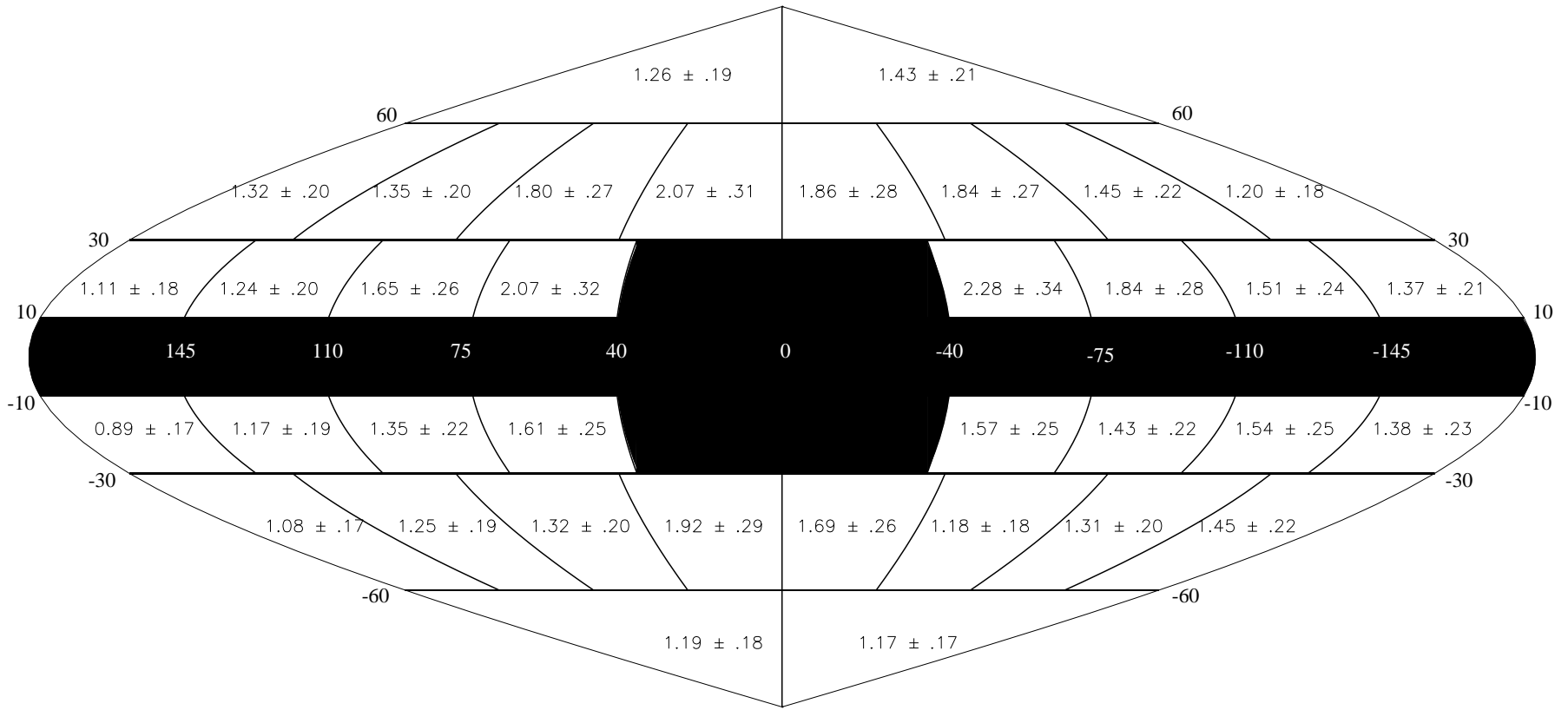
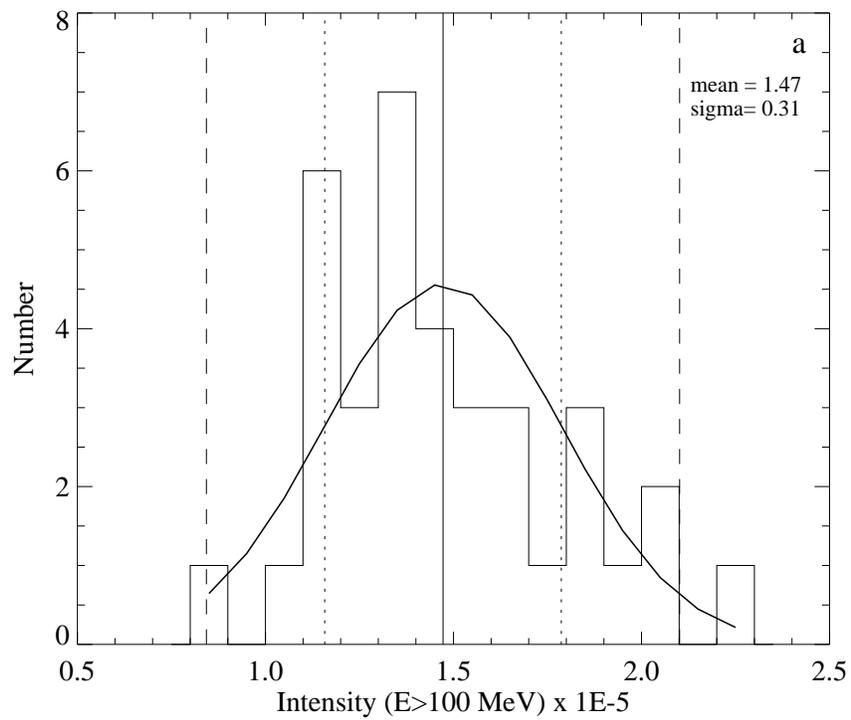
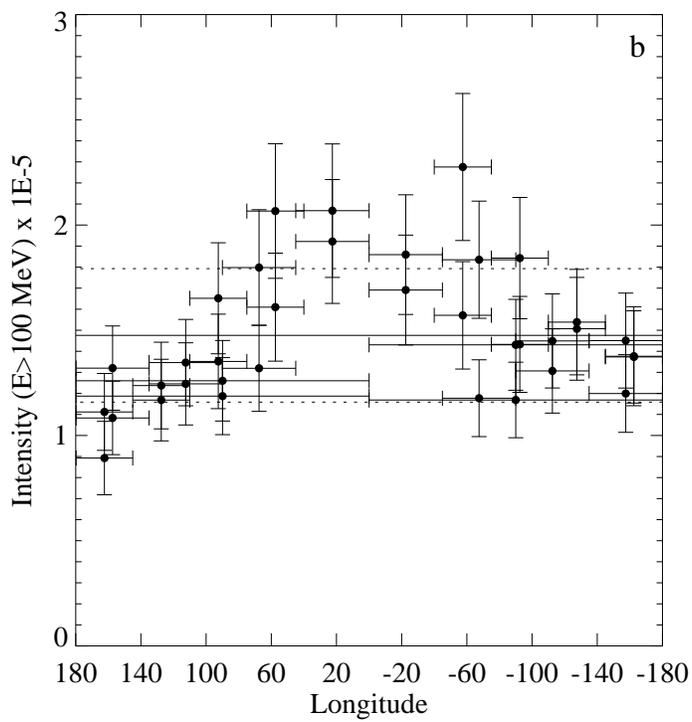
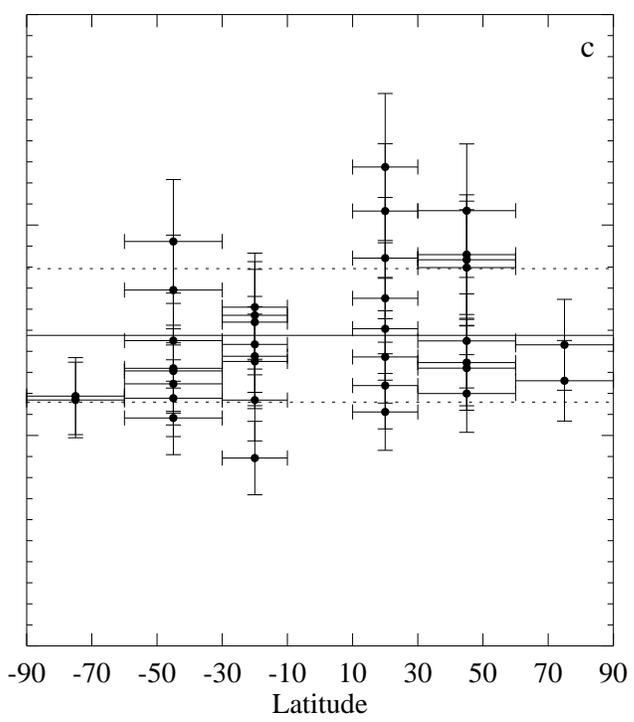
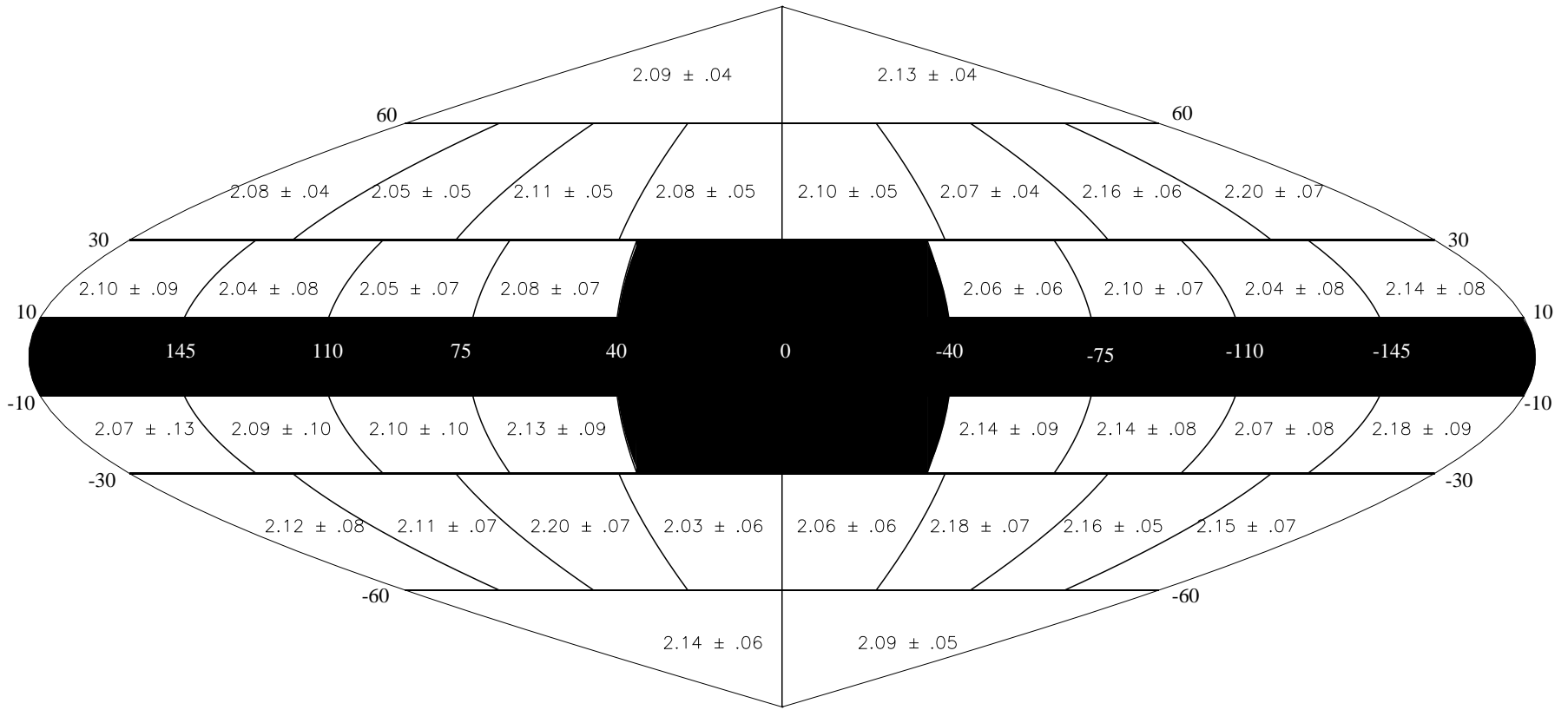
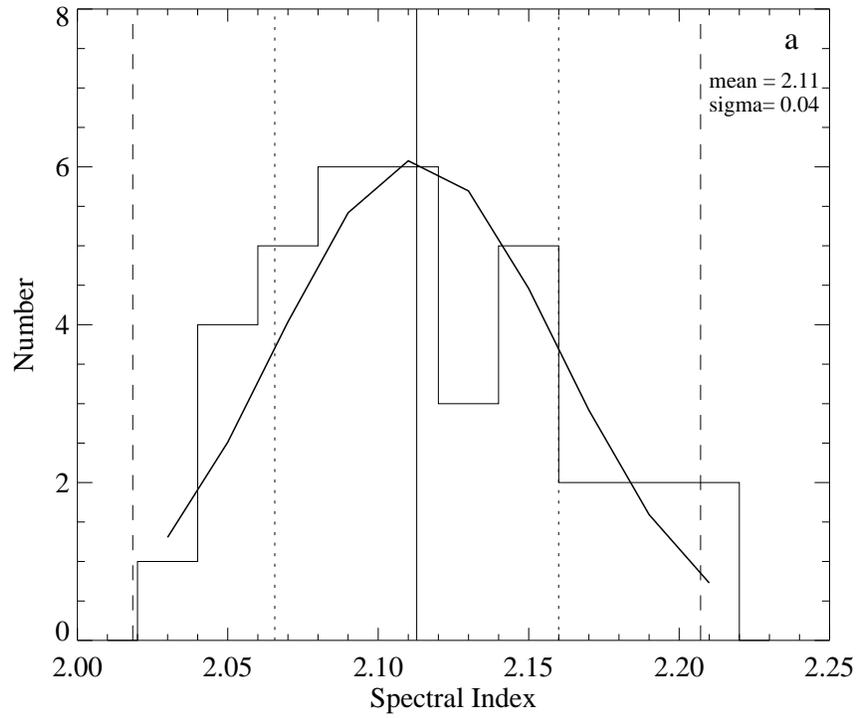
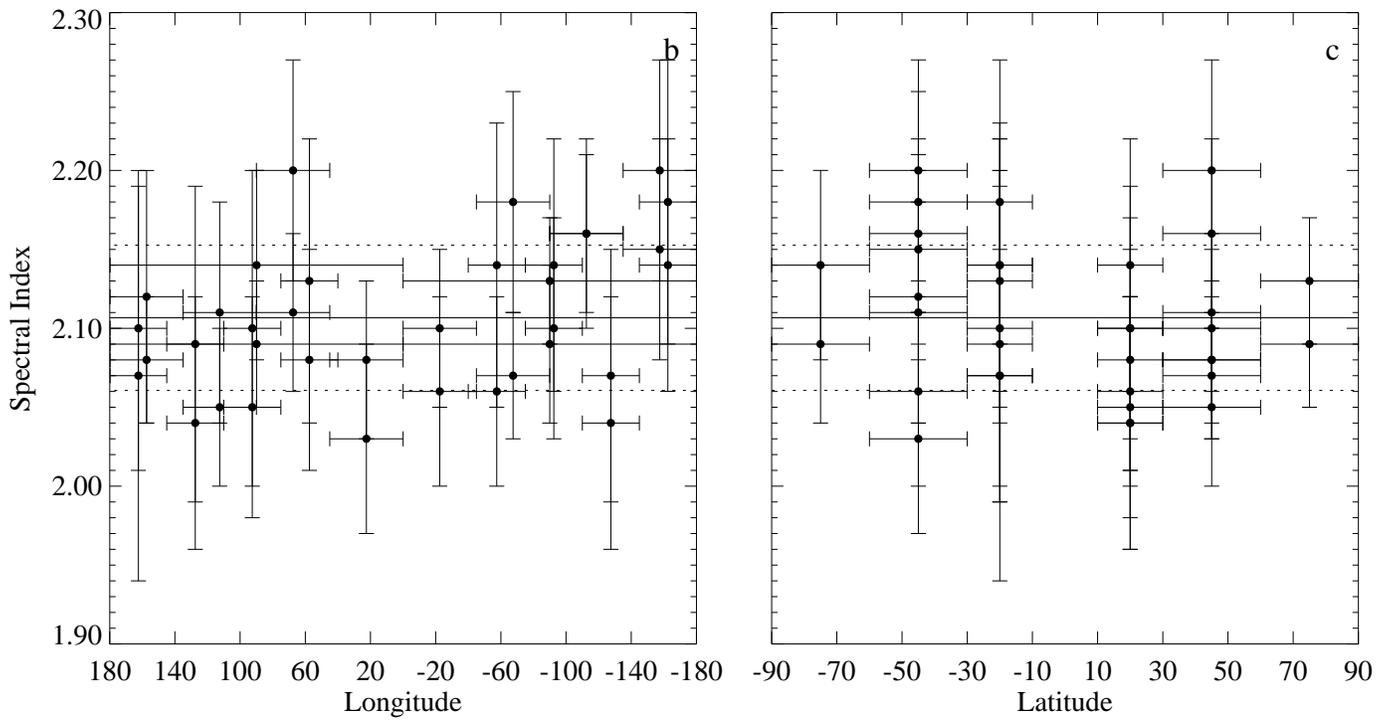
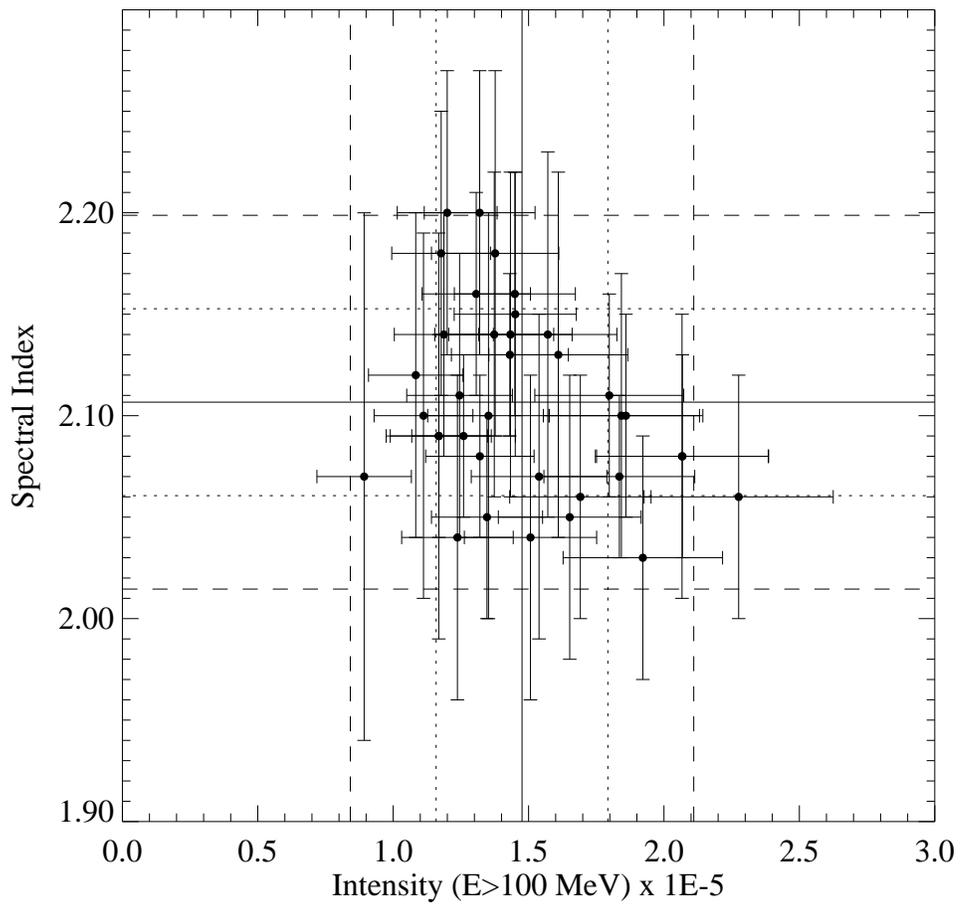
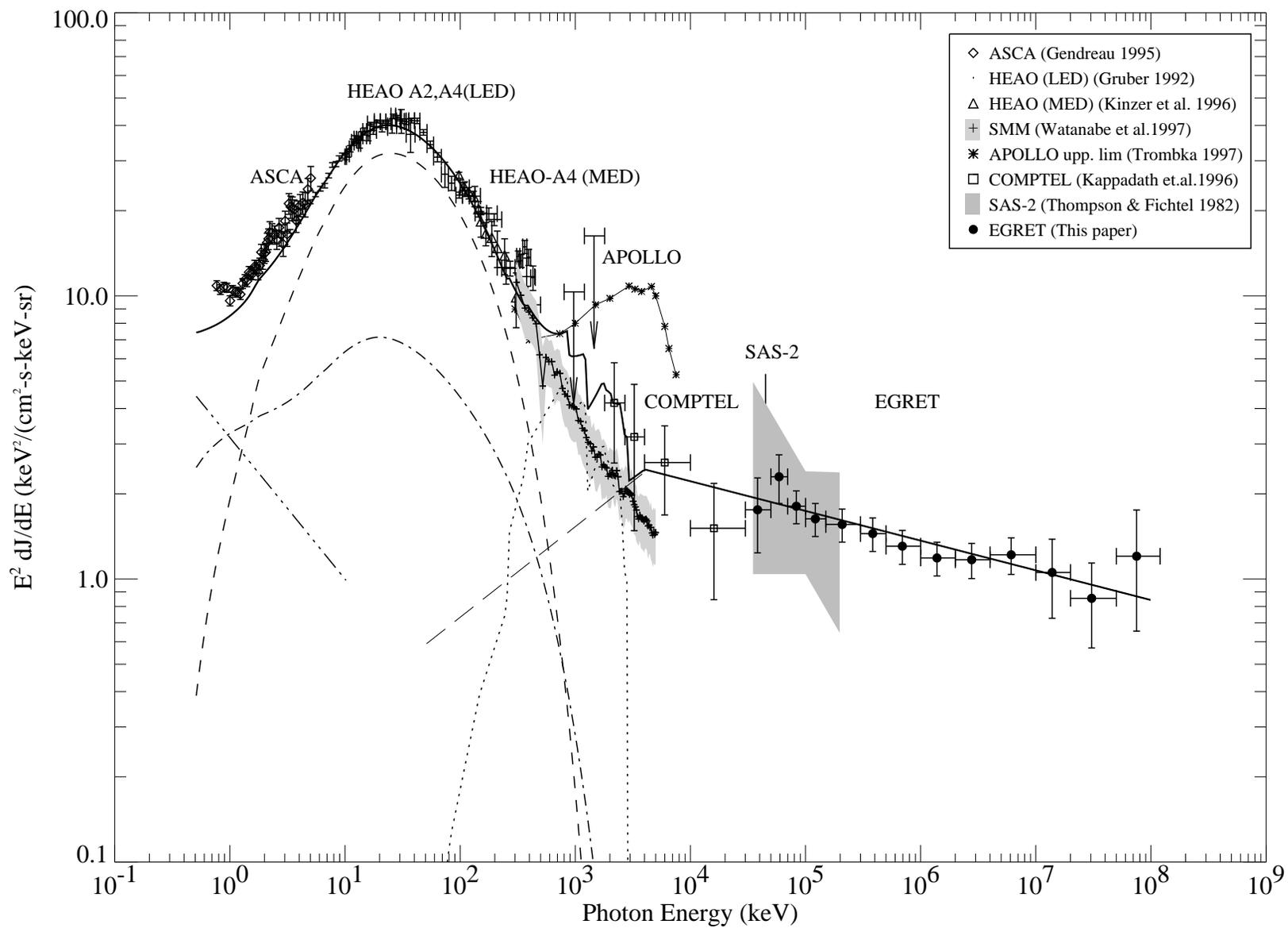
\end{document}